\documentclass[reprint, amsmath, amssymb, preprintnumbers, showpacs, showkeys, aps, prl,floatfix]{revtex4-1}

\usepackage{graphicx}
\usepackage{float}
\usepackage{color}
\begin{document}
\preprint{HAXPES-MCDAD}

\title{Magnetic dichroism in angle-resolved hard x-ray photoemission from
    buried layers.}

\author{Xeniya Kozina}
\author{Gerhard H. Fecher}
\email{fecher@uni-mainz.de}
\author{Gregory Stryganyuk}
\author{Siham Ouardi}
\author{Benjamin Balke}
\author{Claudia Felser}
\affiliation{Institut f\"ur Anorganische und Analytische Chemie,
             Johannes Gutenberg - Universit\"at, 55099 Mainz, Germany.}

\author{Gerd Sch{\"o}nhense}
\affiliation{Institut f\"ur Physik,
             Johannes Gutenberg - Universit\"at, 55099 Mainz, Germany.}

\author{Eiji Ikenaga, Takeharu Sugiyama, Naomi Kawamura, Motohiro Suzuki}
\affiliation{Japan Synchrotron Radiation Research Institute,
             SPring-8, Hyogo, 679-5198, Japan}

\author{Tomoyuki Taira, Tetsuya Uemura, Masafumi Yamamoto}
\affiliation{Division of Electronics for Informatics,
             Hokkaido University, Sapporo 060-0814, Japan}
	
\author{Hiroaki Sukegawa, Wenhong Wang, Koichiro Inomata}
\affiliation{National Institute for Materials Science,
             Tsukuba 305-0047, Japan}

\author{Keisuke Kobayashi}
\affiliation{National Institute for Materials Science,
             SPring-8, Hyogo 679-5148, Japan.}

\date{\today}

\begin{abstract}

This work reports the measurement of magnetic dichroism in angular-resolved
photoemission from in-plane magnetized buried thin films. The high bulk
sensitivity of hard X-ray photoelectron spectroscopy (HAXPES) in combination
with circularly polarized radiation enables the investigation of the magnetic
properties of buried layers. HAXPES experiments with an excitation energy of
8~keV were performed on exchange-biased magnetic layers covered by thin oxide
films. Two types of structures were investigated with the IrMn exchange-biasing
layer either above or below the ferromagnetic layer: one with a CoFe layer on
top and another with a Co$_2$FeAl layer buried beneath the IrMn layer. A
pronounced magnetic dichroism is found in the Co and Fe $2p$ states of both
materials. The localization of the magnetic moments at the Fe site conditioning
the peculiar characteristics of the Co$_2$FeAl Heusler compound, predicted to be
a half-metallic ferromagnet, is revealed from the magnetic dichroism detected
in the Fe $2p$ states.

\end{abstract}


\maketitle
%%%%%%%%%%%%%%%%%%%%%%%%%%%%%%%%%%%%%%%%%%%%%%%%%%%%%%%%
%\section{Introduction}
%%%%%%%%%%%%%%%%%%%%%%%%%%%%%%%%%%%%%%%%%%%%%%%%%%%%%%%%

Rapid breakthroughs in the area of spintronics have led to the development of
electronic devices with improved performance. Being a principal constituent part
of such devices, complex multilayer structures have caused considerable interest
in exploring their unique properties and at the same time have made this task
rather sophisticated. Along with investigation of micromagnetic properties, an
improved understanding of magnetoelectronic properties of deeply buried layers
and interfaces in magnetic multilayer structures is of the most importance in
the viewpoint of their potential applications in the field of magnetic
recording, as data storage devices and sensors.

Magnetic circular dichroism (MCD) in photoabsorption and photoemission has
become a very powerful tool for the element-specific investigation of the
magnetic properties of alloys and compounds. Thus far, such studies have been
mainly carried out using soft X-rays, resulting in a rather surface sensitive
technique due to the low electron mean free path of the resulting low energy
electrons. The application of hard X-rays~\cite{Lind74} results in the emission
of electrons with high kinetic energies and thus, it increases the probing
depth~\cite{Koba03}. The bulk sensitivity of this technique was recently proved
and for $h\nu>8$~keV, the bulk spectral weight was found to reach more than
95\%~\cite{Suga09}. Hard X-ray photoelectron spectroscopy (HAXPES) has been
found to be a well-adaptable non-destructive technique for the analysis of
chemical and electronic states~\cite{FBG08,Koz10}. It was
recently shown that HAXPES can be combined easily with variable photon
polarization when using phase retarders~\cite{Ueda08}. Linear dichroism in the
angular distribution of the photoelectrons is achieved using linearly polarized
hard X-rays and is succesfully applied to identify the symmetry of valence band
states in Heusler compounds~\cite{Ouardi11}. In combination with excitation by
circularly polarized X-rays~\cite{Ueda08}, this method will serve as a unique
tool for the investigation of the electronic and magnetic structure of deeply
buried layers and interfaces.

Baumgarten et al.~\cite{Baum90} carried out a pioneering study on magnetic
dichroism in photoemission and observed this phenomenon in the core-level
spectra of transition metals. The effect, however, was rather small (few {\%})
because of the limited resolution of the experiment. It was later shown that
dichroic effects are also obtained using linearly or even unpolarized
photons~\cite{Ven93,Getz94}. The observed intensity differences in photoemission
are essentially a phenomenon specific to angular-resolved measurements, and
therefore, these have been termed as magnetic circular dichroism in the angular
distribution (MCDAD)~\cite{Cher95,VdL95,Hill96}.

%%%%%%%%%%%%%%%%%%%%%%%%%%%%%%%%%%%%%%%%%%%%%%%%%%%%%%%%%%%%%%%%%%%%%%%%%%%%%%%
\section{Magnetic dichroism in the angular distribution of photoelectrons MDAD.}

Theoretical atomic single-particle models were quite successful in describing,
explaining, and predicting many aspects of magnetic dichroism. Cherepkov et al.
elaborated the general formalism for the dichroism in photoemission excited by
circularly, linearly, and unpolarized radiation~\cite{Cher95}. They showed that
MCDAD is very sensitive to the geometry of the experiment and depends strongly
on the relative orientation between the magnetization, helicity, and momentum of
the excited electrons. The maximum effect is obtained when the magnetization and
helicity vectors are parallel; the effect decreases with an increase in the
angle between these vectors.

The electronic states in solids usually do not carry a spherical or axial
symmetry as in free atoms but have to follow the symmetry of the
crystal~\cite{FKC02}. The angular distribution $I^j({\mathbf{k}},{\bf{n}})$ of
the photoemitted electrons - as derived for example in Reference~\cite{Cher95}
for the case of axially symmetric polarized atoms - has to  account for the non-
diagonal density matrix $\rho_{NM_N'}^{\bf{n}}$~\cite{BLU81}. This leads to the
following equation for the case of a non-axial symmetry:

\begin{widetext}
\begin{equation}
  \begin{array}{ll}
	I^j ( {\mathbf{k}},{\bf{n}} ) & = \dfrac{c_\sigma }{[l]}\sqrt{\dfrac{3\left[ j\right] }{4\pi}}
                              \sum\limits_{\kappa ,L}
                              \sum\limits_N\left[ N \right]^{1/2} C_{\kappa LN}^j
                              \sum\limits_{x,M}
                              \sum\limits_{M_N,M_N'}
                              \rho_{\kappa x}^\gamma
                              \rho_{NM_N'}^{\bf{n}}(j) Y_{LM}^{*}(\mathbf{k}) \\
                          &   \times D_{M_NM'_N}^N(\Omega)
                              \left(
                              \begin{array}{ccc}
	                               \kappa & L & N \\
	                               x      & M & M_N
                              \end{array}
                              \right)
  \end{array}
  \label{eq:pe}
\end{equation}
\end{widetext}

\begin{figure}[htb]
  \includegraphics[width=6cm]{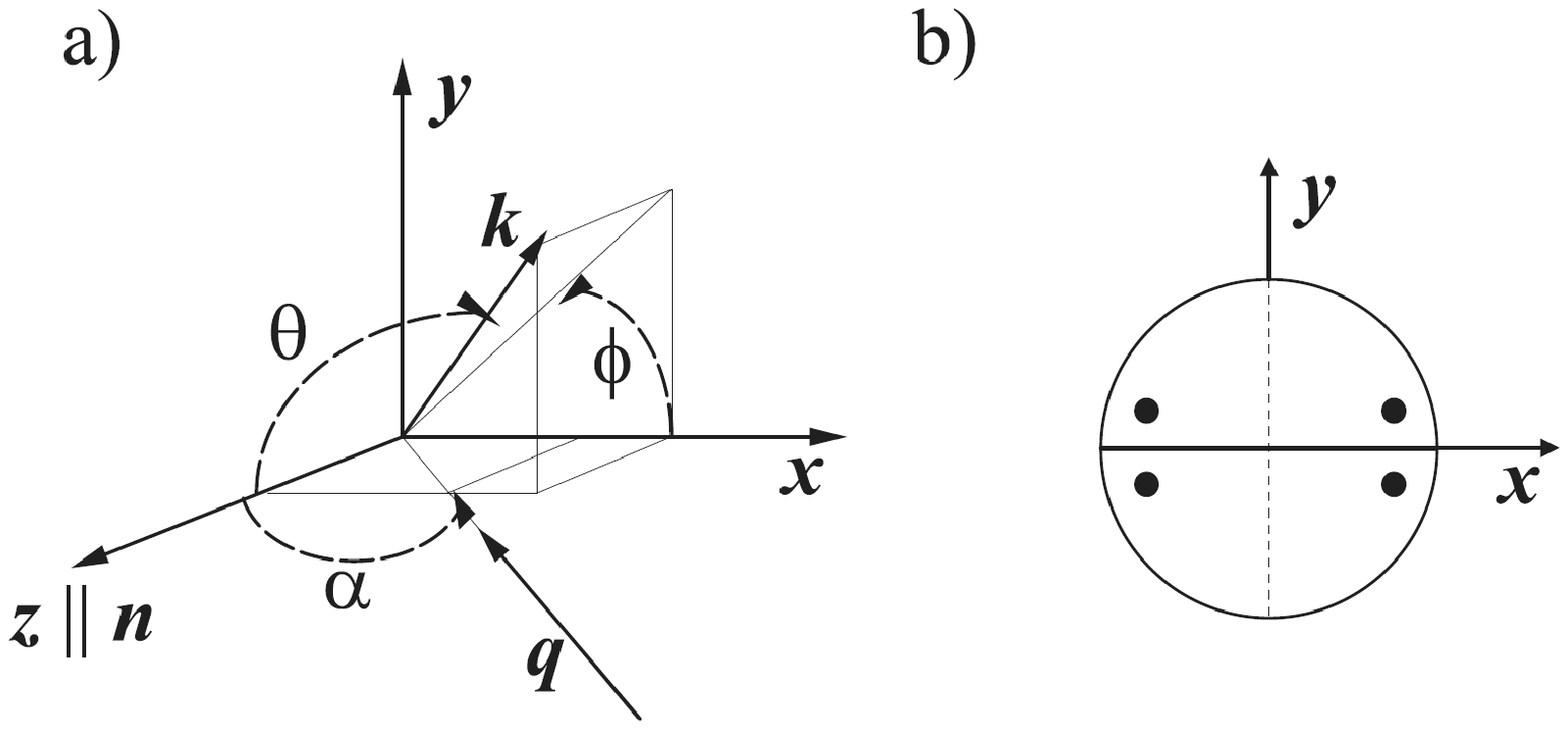}
  \caption{(a)~The coordinate system used for the investigation of photoemission.
           ${\bf k}(\theta,\phi)$ is the electron momentum, $q$ is the photon beam and $n$ is the principal axis of alignment.
           $\theta$ and $\phi$ are the angles defining the direction of the outgoing photoelectrons.
           $\alpha$ is the angle of photon incidence (in the XZ plane) as defined in optics.
           It is seen that the angle describing the photon propagation in spherical coordinates is given by $\Theta_q=\alpha+\pi$.
           The direction of the $z$ axis corresponds to the quantization axis ${\bf n}$.
           (b)~The direction of the in-plane axes $x$ and $y$ is illustrated for an object with $C_{2v}$ symmetry.
           }
\label{fig:vect}
\end{figure}

$l$ and $j$ are the orbital and the total angular momentum of an electron in the
initial state. $C_{\kappa LN}^j$ are the dynamic parameters derived from the
radial matrix elements and $\rho _{\kappa x}^\gamma $ are photon state
multipoles~\cite{BLU81}. $D_{mm_j}^j(\Omega)$ is the Wigner rotation matrix with
$\Omega$ being the set of Euler angles describing the rotation from the
laboratory to the atomic coordinate frame. The direction of the electron
momentum $\stackrel{\rightarrow}{k} ={\bf k}(\theta,\phi)$ is defined by the
angles $\theta$ and $\phi$ (see Figure~\ref{fig:vect}). Finally, $c_\sigma$ is
a photon-energy ($h\nu$) dependent constant: $c_\sigma =\frac{4\pi
^2\alpha\ h\nu}{3}$ where $\alpha$ is the fine structure constant.

%%%%%%%%%%%%%%%%%%%%%%%%%%%%%%%%%%%%%%%%%%%%%%%%%%%%%%%%
%%%%%%%%%%%%%%%%%%%%%%%%%%%%%%%%%%%%%%%%%%%%%%%%%%%%%%%%

This formalism can also be used to consider open shell atoms and the multiplets
resulting from the interaction between the core states and the open shell
valence states. In that case, the dynamic parameters $C_{J\kappa LN}^j$ have to
be calculated for the appropriate coupling scheme ($jj$, LSJ, or intermediate)
with the single particle quantum numbers $j,m$ being replaced by those ($J,M$)
describing the complete atomic state~\cite{Cher95}. In that case, the dynamic
parameter will redistribute the single-electron results in a particular way over
the states of a multiplet (see
References~\cite{TL91a,*TholVdL91,*TL94a,*TholVdL94,Fec01}).

The state multipoles of the $s$, $p$, and $d$-states that define the intensity
and the sign and magnitude of the dichroism are summarized in
Tables~\ref{tab:statemultipoles01} and~\ref{tab:statemultipoles2}. Note that the
state multipoles are independent of the orbital angular momentum $L$, they
depend only on the total angular momentum $J$ and its projection $M_J$.

% Table 1 %%%%%%%%%%%%%%%%%%%%%%%%%%%%%%%%%%%%%%%%%%%%%%%%%%%
\begin{table}[htb]
\centering
\caption{State multipoles of $\left|L, J\right> = \left|0, 1/2\right>$,
         $\left|1, 1/2\right>$, $\left|1, 3/2\right>$, and $\left|2, 3/2\right>$ states.}
    \begin{ruledtabular}
    \begin{tabular}{l|cc|cccc}
        $J$          & $\frac{1}{2}$        &                       & $\frac{3}{2}$ & & & \\
        $M_J$        & $+\frac{1}{2}$       & $-\frac{1}{2}$        & $+\frac{3}{2}$ & $+\frac{1}{2}$ & $-\frac{1}{2}$ & $-\frac{3}{2}$\\
        \noalign{\smallskip}\hline\noalign{\smallskip}
        $\rho_{00}$  & $\frac{1}{\sqrt{2}}$ & $\frac{1}{\sqrt{2}}$  & $\frac{1}{2}$ & $\frac{1}{2}$  & $\frac{1}{2}$  & $\frac{1}{2}$ \\
        $\rho_{10}$  & $\frac{1}{\sqrt{2}}$ & $-\frac{1}{\sqrt{2}}$ & $\frac{3}{2\sqrt{5}}$ & $\frac{1}{2\sqrt{5}}$ & $-\frac{1}{2\sqrt{5}}$ &  $-\frac{3}{2\sqrt{5}}$ \\
        $\rho_{20}$  & -                    & -                     & $\frac{1}{2}$ & $-\frac{1}{2}$ & $-\frac{1}{2}$ & $\frac{1}{2}$ \\
        $\rho_{30}$  & -                    & -                     & $\frac{1}{2\sqrt{5}}$ & $-\frac{3}{2\sqrt{5}}$ & $\frac{3}{2\sqrt{5}}$ &  $-\frac{1}{2\sqrt{5}}$ \\
    \end{tabular}
    \end{ruledtabular}
    \label{tab:statemultipoles01}
\end{table}
%%%%%%%%%%%%%%%%%%%%%%%%%%%%%%%%%%%%%%%%%%%%%%%%%%%%%%%%%%%%%

% Table 2 %%%%%%%%%%%%%%%%%%%%%%%%%%%%%%%%%%%%%%%%%%%%%%%%%%%
\begin{table}[htb]
\centering
\caption{State multipoles of $\left|L, J\right> = \left|2, 5/2\right>$ states.}
    \begin{ruledtabular}
    \begin{tabular}{l|cccccc}
    $J$          & $\frac{5}{2}$        & &  & & & \\
    $M_J$        & $-\frac{5}{2}$       & $-\frac{3}{2}$        & $-\frac{1}{2}$ & $+\frac{1}{2}$ & $+\frac{3}{2}$ & $+\frac{5}{2}$ \\
    \noalign{\smallskip}\hline\noalign{\smallskip}
    $\rho_{00}$  & $\frac{1}{\sqrt{6}}$ & $\frac{1}{\sqrt{6}}$  & $\frac{1}{\sqrt{6}}$ & $\frac{1}{\sqrt{6}}$  & $\frac{1}{\sqrt{6}}$  & $\frac{1}{\sqrt{6}}$ \\
    $\rho_{10}$  & $-\frac{5}{\sqrt{70}}$ & $-\frac{3}{\sqrt{70}}$ & $-\frac{1}{\sqrt{70}}$ & $\frac{1}{\sqrt{70}}$ & $\frac{3}{\sqrt{70}}$ &  $\frac{5}{\sqrt{70}}$ \\
    $\rho_{20}$  & $\frac{5}{2\sqrt{21}}$ & $-\frac{1}{2\sqrt{21}}$ & $-\frac{2}{\sqrt{21}}$ & $-\frac{2}{\sqrt{21}}$ & $-\frac{1}{2\sqrt{21}}$ & $\frac{5}{2\sqrt{21}}$ \\
    $\rho_{30}$  & $-\frac{5}{6\sqrt{5}}$ & $\frac{7}{6\sqrt{5}}$ & $\frac{2}{3\sqrt{5}}$ & $-\frac{2}{3\sqrt{5}}$ & $-\frac{7}{6\sqrt{5}}$ &  $\frac{5}{6\sqrt{5}}$ \\
    $\rho_{40}$  & $\frac{1}{2\sqrt{7}}$ & $-\frac{3}{2\sqrt{7}}$ & $\frac{1}{\sqrt{7}}$ & $\frac{1}{\sqrt{7}}$ & $-\frac{3}{2\sqrt{7}}$ &  $\frac{1}{2\sqrt{7}}$ \\
    $\rho_{50}$  & $-\frac{1}{6\sqrt{7}}$ & $\frac{5}{6\sqrt{7}}$ & $-\frac{5}{3\sqrt{7}}$ & $\frac{5}{3\sqrt{7}}$ & $-\frac{5}{6\sqrt{7}}$ &  $\frac{1}{6\sqrt{7}}$ \\
    \end{tabular}
    \end{ruledtabular}
    \label{tab:statemultipoles2}
\end{table}
%%%%%%%%%%%%%%%%%%%%%%%%%%%%%%%%%%%%%%%%%%%%%%%%%%%%%%%%%%%%%

%%%%%%%%%%%%%%%%%%%%%%%%%%%%%%%%%%%%%%%%%%%%%%%%%%%%%%%%%%%%%%%%%%%%%%%%%%%%%%%
\subsection{MDAD Equations for the grazing incidence geometry.}
\label{sec:eqexp}

In the following, let us consider the special case of geometry with the photons
impinging in the $x-z$ plane with unit vector of the photon momentum $\hat{{q}}
= (-\cos(\alpha),-\sin(\alpha),0)$. At such a grazing incidence with
$\alpha=\pi/2$ it becomes $\hat{{q}} = (-1,0,0)$. The electrons are observed in
the direction perpendicular to the photon beam ($\theta=\frac{\pi}{2}-\alpha$)
with the momentum $\hat{{k}} = (-\sin(\theta),0,\cos(\theta))$. At a photon
incidence of $\alpha=\pi/2$ it becomes $\hat{{k}} = (0,0,1)$. (Compare also
Figures~\ref{fig:vect} and~\ref{fig:geometry}.)

Now examine the case: $\stackrel{\rightarrow}{n} \rightarrow -
\stackrel{\rightarrow}{n}$ where the magnetic dichroism emerges from a switching
of the direction of magnetization with the initial direction
$\stackrel{\rightarrow}{n}=(1,0,0)$ that is along the $x$-axis. Applying
Equation~(\ref{eq:pe}) and the state multipoles of
Table~\ref{tab:statemultipoles01} the circular magnetic dichroism in the angular
distribution for $p$-states is given by the equations:

\begin{widetext}
\begin{equation}
  \begin{array}{ll}
     CMDAD^{\sigma+}(p) & = - \rho_{10} \sin(\alpha) (\sqrt{\frac{2}{3}} C_{JkLN}^{(1, 0, 1)}
                            + \sqrt{\frac{1}{15}} C_{JkLN}^{(1, 2, 1)}(1-6\cos^2(\alpha))) \\	

     CMDAD^{\sigma-}(p) & = + \rho_{10} \sin(\alpha) (\sqrt{\frac{2}{3}} C_{JkLN}^{(1, 0, 1)}
                            + \sqrt{\frac{1}{15}} C_{JkLN}^{(1, 2, 1)}(1-6\cos^2(\alpha)))
  \end{array}
\end{equation}
\end{widetext}

The circular magnetic dichroism in the angular distribution (CMDAD) for opposite
helicity of the photons has an opposite sign. The equations for the $p_{1/2}$
and $p_{3/2}$ states are the same. The magnitude differs, however, because of
the differences in the state multipoles $\rho_{10}$ and dynamical parameters
$C_{JkLN}$.

For $\alpha=\pi/2$ the $CMDAD$ of the $p$-states ($J=1/2, 3/2$) becomes simply:

\begin{equation}
   CMDAD^{\sigma\pm}(p_J) = \mp \rho_{10} \left( \sqrt{\frac{2}{3}} C_{JkLN}^{(1, 0, 1)}
                                                + \sqrt{\frac{1}{15}} C_{JkLN}^{(1, 2, 1)} \right)
\label{eq:cmdad_p}
\end{equation}

The linear counterpart LMDAD vanishes in that geometry, independent whether the
photons are $s$ or $p$ polarized. At $\alpha=\pi/2$ the magnetic dichroism in
the angular distribution vanishes for all $p$-states independent of the
polarization of the photons if the magnetization is perpendicular to the plane
spanned by the photon incidence and the electron momentum (here for the x-z
plane with $\stackrel{\rightarrow}{n}=(0,\pm 1,0)$).

%%%%%%%%%%%%%%%%%%%%%%%%%%%%%%%%%%%%%%%%%%%%%%%%%%%%%%%%
\section{Experimental details}
%%%%%%%%%%%%%%%%%%%%%%%%%%%%%%%%%%%%%%%%%%%%%%%%%%%%%%%%

\begin{figure}[htb]
  \includegraphics[width=7.5cm, clip]{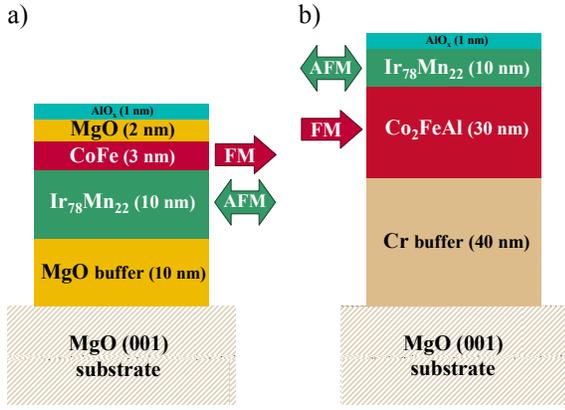}
  \caption{(Color online) Sketch of the exchange-biased films used in the dichroism experiments.
           The multilayer structure in (a) corresponds to the lower part of the electrode
           and is realized in on-top configuration with CoFe ferromagnetic layer. The structure
           shown in (b) presents on-bottom configuration with Co$_2$FeAl ferromagnetic layer.
           In both films a 1-nm-thick AlO$_x$ layer is used as a protective cap.
           }
\label{fig:samples}
\end{figure}

The present study reports on the MCDAD experiment in the HAXPES range on
different types of exchange-biased structures with epitaxially grown
ferromagnetic layers of CoFe and Co$_2$FeAl, these being typical materials used
in tunnel magnetoresistive devices. (see Figure~\ref{fig:samples}) The on-top
approach multilayers were deposited in the sequence MgO(100) substrate~/ MgO
buffer layer (10\,nm)~/ Ir$_{78}$Mn$_{22}$ (10\,nm)~/ CoFe (3\,nm)~/ MgO barrier
(2\,nm)~/ AlO$_x$ (1\,nm)~\cite{Mas08} that corresponds to the lower
exchange-biased electrode of a magnetic tunnel junction (MTJ). After growth the stacks
were annealed at $350^{\circ}$C for 1~h in vacuum of $5\times 10^{-2}$~Pa in a
magnetic field of 0.4~MAm$^{-1}$ to provide exchange biasing of the CoFe layer film
through the IrMn/CoFe interface (see also~\cite{Mar07}). The on-bottom
configuration was realized in the multilayer sequence MgO(100) substrate~/ Cr
buffer layer (40~nm)~/ Co$_2$FeAl (30~nm)~/ Ir$_{78}$Mn$_{22}$ (10~nm)~/ AlO$_x$
(1~nm)~\cite{Wang09}. The sample stacks were annealed at $400^{\circ}$C for 1~h
in vacuum under a magnetic field of 0.4~MAm$^{-1}$ to provide exchange biasing
to the Co$_2$FeAl thin film through the Co$_2$FeAl/IrMn interface (see
also~\cite{Wang09}). In both cases, the topmost AlO$_x$ layers served as a
protective coating. All metal layers were deposited by magnetron sputtering and
electron beam evaporation was used to epitaxially grow the MgO barrier. IrMn
serves as an exchange-biasing layer that keeps CoFe or Co$_2$FeAl magnetized in
preset directions.

%%%%%%%%%%%%%%%%%%%%%%%%%%%%%%%%%%%%%%%%%%%%%%%%%%%%%%%%
%%%%%%%%%%%%%%%%%%%%%%%%%%%%%%%%%%%%%%%%%%%%%%%%%%%%%%%%

\begin{figure}[htb]
  \includegraphics[width=7.5cm, clip]{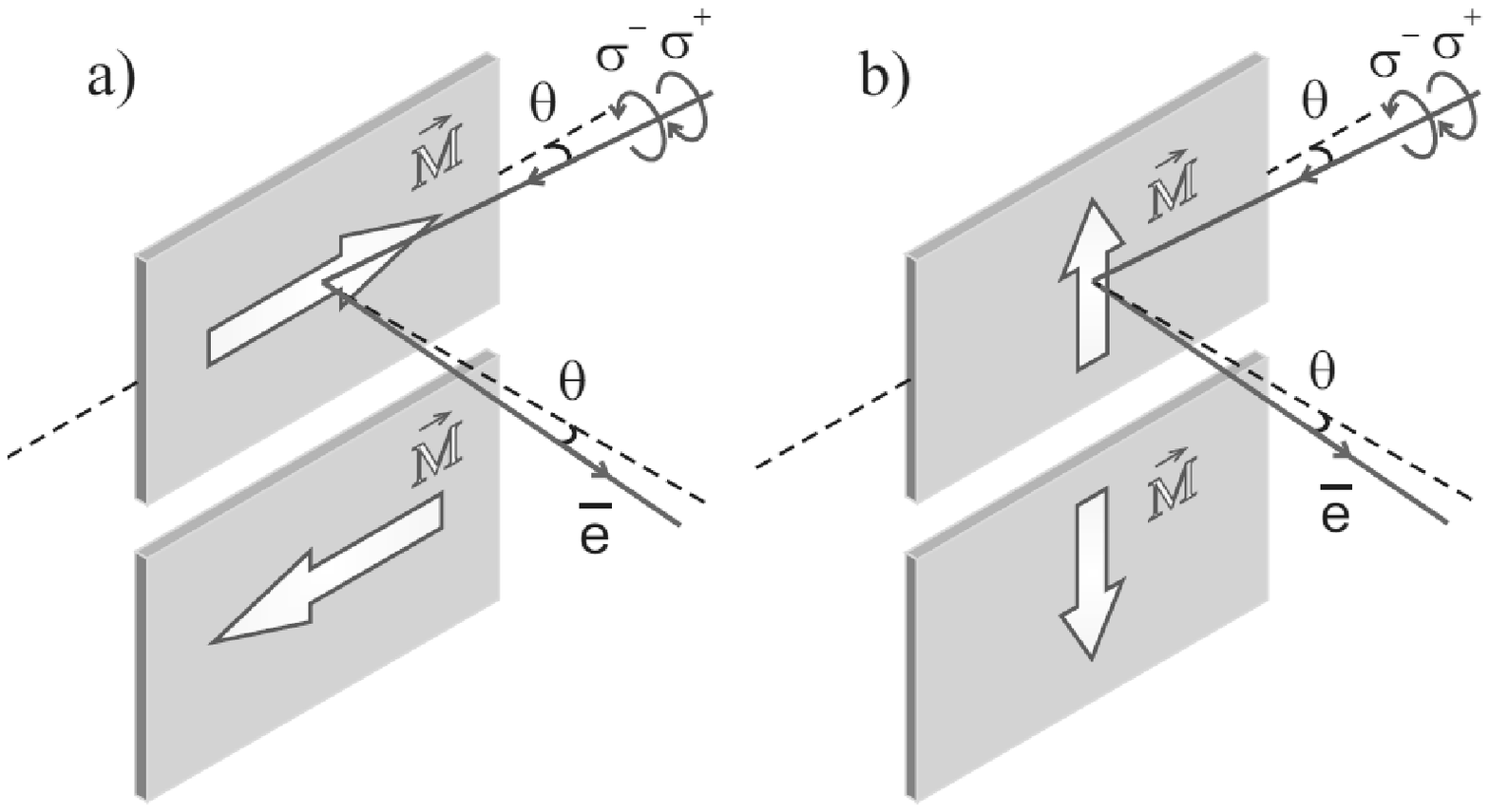}
  \caption{(Color online) Scheme of the experimental geometry. \newline
          The incidence angle $\theta$ (with respect to the surface plane)
          of the circularly polarized photons was fixed to $2^{\circ}$.
          X-rays of opposite helicity ($\sigma^+$ and $\sigma^-$) were provided by a
          phase retarder. Further, samples with opposite directions of magnetization are used.
          In (a) the in-plane magnetization ${\textbf{M}}$ is nearly parallel to the beam axis and
          in (b) the in-plane magnetization is perpendicular to the beam axis. The electron detection
          is fixed and perpendicular to the photon beam.
           }
\label{fig:geometry}
\end{figure}

The magnetized samples were mounted pairwise with opposite magnetization on the
same sampleholder and can be selected via sample shift. Care was taken that the
magnetization directions were antiparallel and that surfaces were parallel to
avoid different detection angles. The mounting of the samples at the fixed
sample manipulator was chosen to have up/down as well as left/right pairs as it
is shown in the Fure~\ref{fig:geometry}). This allowed to probe the dichroism by
varying both the direction of magnetization and the direction of helicity .

%%%%%%%%%%%%%%%%%%%%%%%%%%%%%%%%%%%%%%%%%%%%%%%%%%%%%%%%
%%%%%%%%%%%%%%%%%%%%%%%%%%%%%%%%%%%%%%%%%%%%%%%%%%%%%%%%

The HAXPES experiments with an excitation energy of 7.940~keV were performed
using beamline BL47XU at SPring-8~\cite{Kob09}. The energy distribution of the
photoemitted electrons was analyzed using a hemispherical analyzer (VG-Scienta
R4000-12kV) with an overall energy resolution of 150~meV or 250~meV. The angle
between the electron spectrometer and the photon propagation was fixed at
$90^{\circ}$. The detection angle was set to $\theta=2^{\circ}$ in order to
reach the near-normal emission geometry and to ensure that the polarization
vector of the circularly polarized photons is nearly parallel ($\sigma^-$) or
antiparallel ($\sigma^+$) to the in-plane magnetization $M^+$. The sign of the
magnetization was varied by mounting samples with opposite directions of
magnetization ($M^+$, $M^-$). The polarization of the incident photons was
varied using an in-vacuum phase retarder based on a 600-$\mu$m-thick diamond
crystal with (220) orientation~\cite{SKM98}. The direct beam is linearly
polarized with $P_p=0.99$. Using the phase retarder, the degree of circular
polarization is set such that $P_{c} > 0.9$. The circular dichroism is
characterized by an asymmetry that is defined as the ratio of the difference
between the intensities $I^+$ and $I^-$ and their sum, $A=(I^+- I^-)/(I^++I^-)$,
where $I^+$ corresponds to $\sigma^+$- and $I^-$ to $\sigma^-$- type helicity.
Magnetic dichroism may be defined in a similar manner using the differences in
the intensities if the direction of the magnetization is changed keeping the
polarization of the photons fixed.

The photon flux on the sample was about $10^{11}$ photons per second in a
bandwidth of $10^{-5}$ during the measurements at the given excitation energy.
The vertical spot size on the sample is  30~$\mu$m, while in horizontal
direction, along the entrance slit of the analyzer, the spot was stretched to
approximately 7~mm. The measurements were performed using grazing incidence
geometry. The resulting count rates (taken from the equivalent gray scale values
provided by the spectrometer software) were in the order of 0.6~to~6~MHz for the
core level spectra including shallow core levels and about 0.25~MHz for the
valence band.

%%%%%%%%%%%%%%%%%%%%%%%%%%%%%%%%%%%%%%%%%%%%%%%%%%%%%%%%
\section{Results and Discussion}
%%%%%%%%%%%%%%%%%%%%%%%%%%%%%%%%%%%%%%%%%%%%%%%%%%%%%%%%

\begin{figure}
  \centering
  \includegraphics[width=5cm]{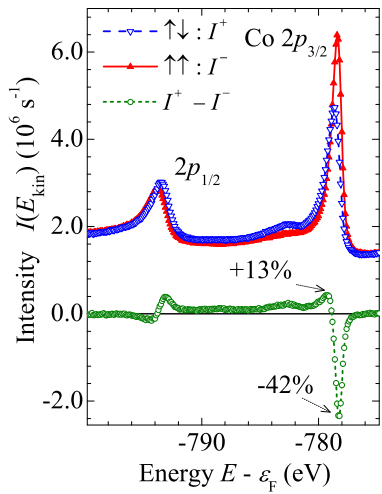}
  \caption{(Color online) Polarization-dependent photoelectron spectra of the Co $2p$ core-level emission from CoFe
           on top of an IrMn exchange-biasing layer and the difference of two spectra.
           Asymmetry values are marked at selected energies. }
\label{fig_Co2p}
\end{figure}

Figure~\ref{fig_Co2p} shows the $2p$ core-level spectra of Co that were taken
from an exchange-biased CoFe film that was covered by oxide films. Pronounced
difference was observed in the spectra taken with photons having opposite
helicity for a fixed direction of magnetization. The pure difference $\Delta
I=I^+ - I^-$ presented in the figure is already free of the influence of the
background and gives the correct shape of the magnetic dichroism. That means, it
contains all characteristic features of the magnetic dichroism. For
quantification and comparison of the dichroic effects, the MCDAD asymmetry was
determined from

\begin{equation}
 A = \frac{(I^+ - I^-)}{(I^+ + I^-)} = \frac{\Delta I}{2 I}
\end{equation}

after subtracting a Shirley-type background from the spectra to find the
asymmetry caused only by the direct transition. The background subtraction
leads, however, to a very low intensity in the beginning, in the end of the
spectral energy range as well as in the range between the spin-orbit split peaks
in both spectra (that is in the ranges of the spectra where no signal from the
transition itself is expected). This, in turn, leads to very high and rather
unphysical values of the calculated asymmetry in these energy ranges. From the
above remark on $\Delta I$ it is therefore advantageous to show the differences
of the intensities and to mark the asymmetry for characteristic energies only.
Here the largest obtained asymmetry value is -42\% at Co $2p_{3/2}$.

%%%%%%%%%%%%%%%%%%%%%%%%%%%%%%%%%%%%%%%%%%%%%%%%%%%%%%%%
%%%%%%%%%%%%%%%%%%%%%%%%%%%%%%%%%%%%%%%%%%%%%%%%%%%%%%%%

As one can see, the spin-orbit splitting of the Co $2p$ states is clearly
resolved, as expected. When going from $p_{3/2}$ to $p_{1/2}$, the dichroism
changes its sign across the $2p$ spectra in the sequence: $-\:+\:\:+\:-$; as
appears characteristic of a Zeemann-type $m_j$ sub-level ordering.
This sequence of signs is directly expected from
Equation~(\ref{eq:cmdad_p}) and the state multipoles $\rho_{10}$ given in
Table~\ref{tab:statemultipoles01} when identifying the states of the
magnetically split $2p$ doublet as $\left|j,m_j\right\rangle$ in the single
particle description. The details of the MCDAD reveal, however, that the
situation is more complicated. In particular, the dichroism in the Fe $2p$
spectra does not vanish in the region between the spin-orbit doublet.
The multiplet formalism to describe the spectra in more detail
will be given below.

MCDAD has previously been used to investigate the itinerant magnetism of
ferromagnetic elements such as Co, Fe, and Ni, where it was explained in terms
of single-particle models~\cite{Hill96,Tob98,Henk99,Bans99}. As demonstrated in
the case of Ni, however, the single-particle approach poorly describes all the
peculiarities of the complex spectra. van der Laan and Thole considered the
MCDAD phenomenon by taking into account the influence of electron correlation
effects in the frame of atomic many-particle models that were successfully used
to describe both localized and itinerant magnetism
phenomena~\cite{TholVdL91,*TholVdL93,*TholVdL94,*VdL95}. Many-body effects play
an important role when using polarized incident photons. The correlation between
spin and orbital moments, $2p$ core-hole, and spin-polarized valence band
results in a rich multiplet structure that spreads out over a wide energy range
of a spectrum~\cite{VdL00}. In strongly correlated systems, the bulk magnetic
and electronic properties are quite different from the surface ones. However, as
observed previously, MCDAD with radiation in the soft X-ray range is highly
sensitive to the surface where the dichroism is influenced by symmetry
breaking~\cite{Fech95}. Because of the strong inelastic electron scattering in
this energy range, the escape depth of the photoemitted electrons of a few {\AA}
becomes comparable to the thickness of a monolayer. The tuning of the excitation
energy also affects the photoionization cross sections. At high energies, the
intensities from the $d$ states of transition metals are reduced as compared to
the partial cross sections of the $s$ and $p$
states~\cite{Koba03,Pana05,Fech07}. The shape and magnitude of the asymmetry
depend on the partial bulk to surface spectral weights; hence, only at high
energies, the dichroism effects appear to be related to the bulk properties.

\begin{figure}[htb]
  \includegraphics[width=5cm]{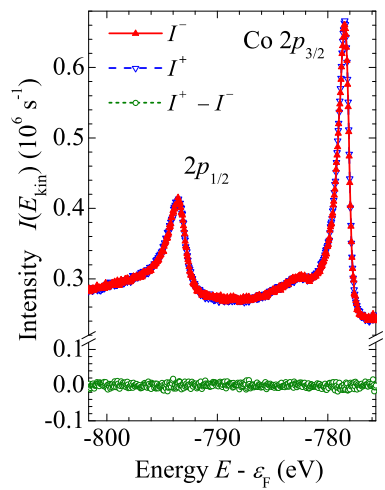}
  \caption{(Color online) Illustration of the vanishing dichroism in photoemission when the
           photon polarization vector is perpendicular to the in-plane magnetization
           vector demonstrated for the Co $2p$ state of CoFe.
           Shown are the photoelectron spectra $I^+$, $I^-$ and their
           difference $I^+ - I^-$ obtained with different helicity
           at fixed magnetization perpendicular to the photon beam. }
\label{fig:CoFedownCo2p}
\end{figure}

It was carefully proven that the dichroism vanished in the geometry in which the
projection of the photon vector is perpendicular to the magnetization,
independently of whether the photon helicity or the magnetization was reversed.
This indicates that the films are perfectly magnetized in the direction forced
by the exchange-biasing layer magnetization. As an example,
Figure~\ref{fig:CoFedownCo2p} confirms the absence of the dichroic signal at the
Co $2p$ states of the CoFe film in agreement to the theoretical description
given above.

%%%%%%%%%%%%%%%%%%%%%%%%%%%%%%%%%%%%%%%%%%%%%%%%%%%%%%%%
%%%%%%%%%%%%%%%%%%%%%%%%%%%%%%%%%%%%%%%%%%%%%%%%%%%%%%%%

\begin{figure}
  \centering
  \includegraphics[width=5cm]{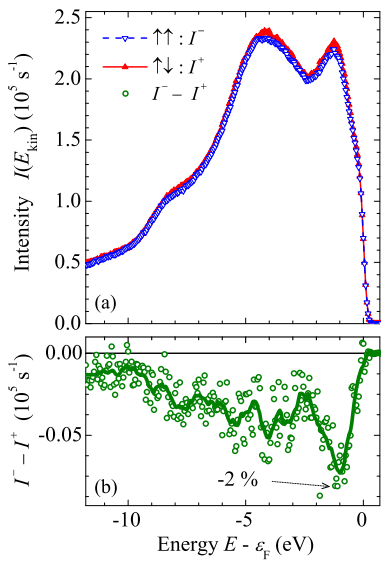}
  \caption{(Color online) MCDAD in valence band of CoFe on top of IrMn.\\
           The asymmetry is given at -1 eV below Fermi level. }
\label{fig_VB}
\end{figure}

Figure~\ref{fig_VB} shows the polarization dependence of the CoFe valence band
spectra together with the resulting magnetic dichroism. The MCDAD observed for
the valence band is much smaller as compared to the core-level photoemission.
The largest asymmetry is approximately -2\% at -1~eV below the Fermi energy.
Such low asymmetry values were also observed when using low photon and kinetic
energies~\cite{Ven97}. Only for excitation close to threshold, higher
asymmetries arise in the case of one-~\cite{Naka06} and two-photon
photoemission~\cite{Hil09}. In the range of the valence states, the detection is
further complicated by the signal from the underlying IrMn layer that does not
contribute to the dichroism. Because of the thin layer of CoFe and the large
escape depth of the nearly 8~keV fast electrons, the two layers cannot be
distinguished in the valence band. It is worthwhile to note that the dichroic
signal itself arises exclusively from the buried, ferromagnetic CoFe layer.

%%%%%%%%%%%%%%%%%%%%%%%%%%%%%%%%%%%%%%%%%%%%%%%%%%%%%%%%
%%%%%%%%%%%%%%%%%%%%%%%%%%%%%%%%%%%%%%%%%%%%%%%%%%%%%%%%

For studies aimed toward the development of novel devices, it is necessary to
also detect the magnetic signal from deeply buried layers. To prove the
reliability of the proposed method, experiments were also performed on samples
in which the IrMn exchange-biasing layer was on top of the layer structure.

\begin{figure}
  \centering
  \includegraphics[width=8cm]{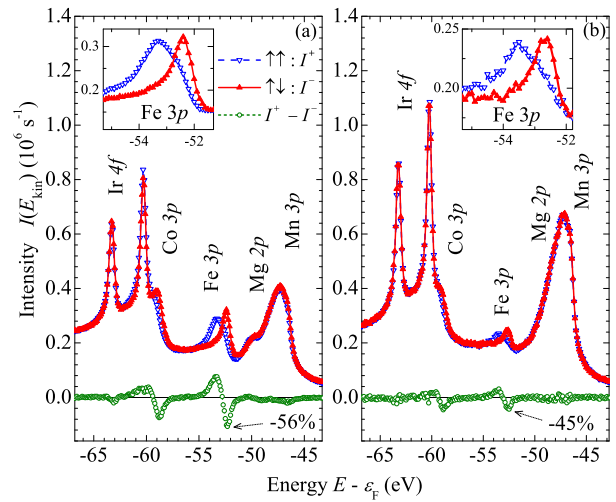}
  \caption{(Color online) MCDAD for the shallow core level spectra obtained from the buried CoFe
           on top and Co$_2$FeAl beneath a 10-nm-thick IrMn film.
           The insets show an enlarged view of  $I^+$ at the Fe $3p$ states. }
\label{fig_SCL}
\end{figure}

Figure~\ref{fig_SCL} compares the MCDAD results for the shallow
core levels of CoFe in the on-top configuration (a) and the deeply buried
Co$_2$FeAl in the on-bottom configuration beneath a 10-nm-thick IrMn film (b).
For such complex multilayer structures, the situation becomes complicated in
that the signals from all the elements contained in the system are detected. In
both cases the shallow core levels of all elements of the multilayers are
detected. The intensity differences between Fe and Co $3p$ emission or Ir $4f$
and Mn $3p$ in the different configurations are obvious and arise from the
damping of the intensity when the electrons pass through the layers above the
emitting layer. Strong signals are still detected from the buried elements even
though the ferromagnetic Co$_2$FeAl layer lies 10~nm beneath the
antiferromagnetic IrMn layer, as it is clearly seen in the inset of
Figure~\ref{fig_SCL}(b). A large asymmetry is clearly observed at the Co and Fe
signals, and these are the ones responsible for the ferromagnetic properties of
the system. The asymmetries of -56\% (CoFe) and -45\% (Co$_2$FeAl) in the Fe
$3p$ signal are quite evident. In Co $3p$, it is well detected even though the
direct spectra overlap with the Ir $4f$ states.

%%%%%%%%%%%%%%%%%%%%%%%%%%%%%%%%%%%%%%%%%%%%%%%%%%%%%%%%
%%%%%%%%%%%%%%%%%%%%%%%%%%%%%%%%%%%%%%%%%%%%%%%%%%%%%%%%

%%%%%%%%%%%%%%%%%%%%%%%%%%%%%%%%%%%%%%%%%%%%%%%%%%%%%%%%
\begin{figure}
  \centering
  \includegraphics[width=8cm]{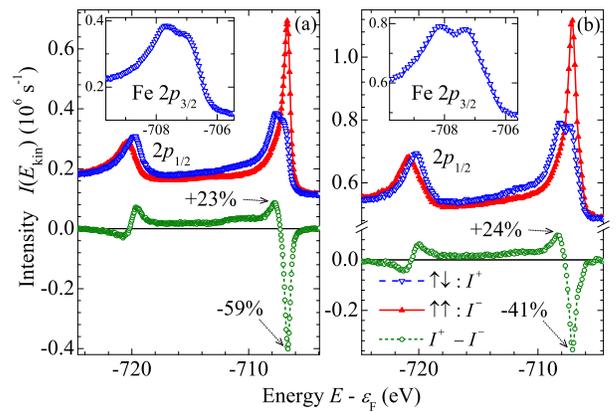}
  \caption{(Color online) Polarization-dependent photoelectron spectra of the Fe $2p$ core-level emission from CoFe
           on top of an IrMn exchange-biasing layer, Co$_2$FeAl beneath IrMn and the corresponding
           differences of the spectra taken with the opposite helicity of light.
           Asymmetry values are marked at selected energies. The insets
           show an enlarged view of $I^+$ at the Fe $2p_{3/2}$ states in both cases.}
\label{fig_CL}
\end{figure}
%%%%%%%%%%%%%%%%%%%%%%%%%%%%%%%%%%%%%%%%%%%%%%%%%%%%%%%%

Figure~\ref{fig_CL} shows the polarization dependent HAXPES
spectra and the MCDAD at the Fe $2p$ states of the buried CoFe (a) and
Co$_2$FeAl (b) layers. The multiplet splitting at the Fe $2p_{3/2}$ is very well
resolved and the MCDAD is well detected in both materials. The emission from the
Co$_2$FeAl has a lower intensity and the resolution was therefore reduced to
250~meV in order to keep the counting rates comparable to those of the CoFe
measurements. (Note that this does not influence the spectra much as they are
governed by a lifetime broadening that is in the same order of magnitude.) It
was shown~\cite{Ros97} that linear magnetic dichroism (LMDAD) along with the
circular one can be successfully applied to investigate the electronic and
magnetic properties of surfaces and interfaces. The LMDAD asymmetry observed at
Fe $2p_{3/2}$, however, was only at most -9\% for a low excitation energy. In
our studies the maximum asymmetries are -59\% for CoFe and - 41\% for Co$_2$FeAl
at Fe $2p_{3/2}$, and this is ideal for the analysis of the magnetic properties.

A closer inspection to the MCD spectra (see insets of Figures~\ref{fig_CL} (a)
and (b)) revealed a striking distinction between the Fe $2p$ spectra of the two
layer systems. Even though taken with a slightly lower resolution, the multiplet
splitting of the Fe $2p_{3/2}$ emission from Co$_2$FeAl appears more pronounced
as compared to the corresponding spectrum from CoFe. The mean splitting $\Delta
E$ of the Fe $2p_{3/2}$ states is 0.8~eV and 1.0~eV for CoFe and Co$_2$FeAl,
respectively. Co$_2$FeAl is supposed to be a half-metallic
ferromagnet with a magnetic moment of 5~$\mu_B$ in the primitive cell and about
2.8~$\mu_B$ per Fe atom~\cite{Wurmehl06}, whereas CoFe is a regular band
ferromagnet with a very high magnetic moment (about 2.5~$\mu_B$ at
Fe)~\cite{Richter88}. In both cases the Fe moment is clearly above that of pure
Fe (2.1~$\mu_B$). One of the major differences is the localized magnetic
moment of Fe in Co$_2$FeAl that is caused by a strong localization of the
$t_{2g}$ bands. In the ordered case of both compounds, the Fe atoms are in a
cubic environment and are surrounded by 8 Co atoms. Co$_2$FeAl forms a perfect
$2^3$ CsCl supercell with every second Fe atom of CoFe replaced by Al. This
causes additional Co-Al bonds that reduce the Co-Fe $d$-state overlap. The
result is a localized moment at the Fe sites. From this viewpoint, Fe in
Co$_2$FeAl is in closer to an covalent than a metallic state. For the Fe atoms,
this causes a more pronounced interaction of the core hole at the ionized $2p$
shell with the partially filled $3d$ valence shell.

%%%%%%%%%%%%%%%%%%%%%%%%%%%%%%%%%%%%%%%%%%%%%%%%%%%%%%%%
\begin{figure}
  \centering
  \includegraphics[width=8cm]{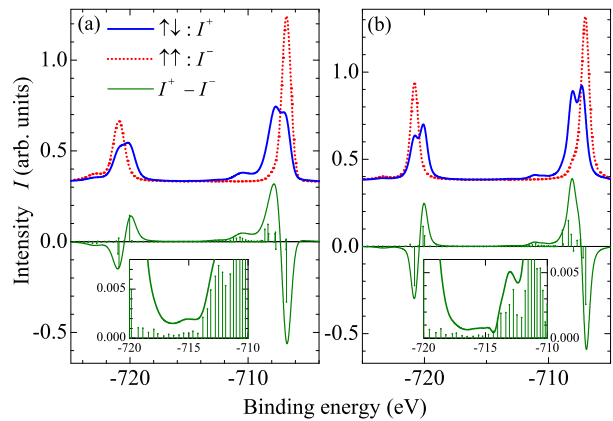}
  \caption{(Color online) Calculated polarization-dependent photoelectron spectra of the
           Fe $2p$ core-level emission obtained by means of atomic multiplet
           calculations and their difference for CoFe (a) and Co$_2$FeAl (b).
           The insets show the enlarged views of the difference curve in the
           region between spin-orbitally splitted components of Fe $2p$ states.
           The bars mark the multiplet states.}
\label{fig_Fe2p_MCD_calc}
\end{figure}
%%%%%%%%%%%%%%%%%%%%%%%%%%%%%%%%%%%%%%%%%%%%%%%%%%%%%%%%

As mentioned above, the single particle theory is not capable
to explain the details of the spectra and their dichroism. It is necessary to
respect the coupling between the ionized core and open valence shells. In the
present case this is the interaction between the $2p^5$ core hole and the open
$3d$ valence shell of Fe. Therefore, multiplet calculations were carried out to
explain the experimentally obtained results for the two different materials.
They were performed by means of CTM4XAS~5.2 (Charge Transfer Multiplet
Calculations for X-ray Absorption Spectroscopy) program~\cite{Stavitski2010},
using its X-ray photoelectron spectroscopy (XPS) option. The results are shown
in Figure~\ref{fig_Fe2p_MCD_calc}. The simulations were made for a Fe$^{\rm 3+}$
ionic ground state with $4s^03d^5$ configuration that describes well the
emission from the Fe-$2p$ states of both systems, CoFe and Co$_2$FeAl. The
Slater integrals ($F\rm_{dd}$, $F\rm_{pd}$, and $G\rm_{pd}$) were reduced to
0.65; 0.55; 0.65 and 0.7; 0.5; 0.5 of the free atom values to describe the
spectra of CoFe and Co$_2$FeAl, respectively. As exchange interaction plays an
important role in ferromagnetic materials, the effect of exchange splitting was
taken into account by setting the magnetic splitting parameter $M$ to 50~meV for
CoFe and 450~meV for Co$_2$FeAl. The obtained values for the splitting $\Delta
E$ of the Fe $2p_{3/2}$ states are 0.9~eV and 1.1~eV for CoFe and Co$_2$FeAl,
respectively. The applied parameters resulted in a quite good agreement between
calculated and experimental spectra and dichroism . Possible, slight
disagreements may be attributed to the fact that the observed spectra depend on
the degree of localization or itineracy of the magnetic moment at the Fe site
through the coupling of the $2p^5$ core hole with the $d$ valence bands.
Fractional $d$ state occupancies (for example $d^{5+x}$, $0<x<1$) that might
better describe the partial delocalization of $d$ electrons of Fe in metallic
systems, however, are not available in the atomic model. The insets in
Figure~\ref{fig_Fe2p_MCD_calc} present a enlarged view of the region of the
dichroism between the main lines of the multiplet. In those insets one clearly
recognizes the appearance of multiplet states over the entire energy range.
These states form the characteristic structure of the dichroism that is in a
good agreement with the experiment.

It is worthwhile to note that such differences between two very
similar alloys are not resolved by X-ray circular dichroism (XMCD) in soft X-ray
photo absorption~\cite{Chen90,*Chen95}. This is found easily if comparing the
here shown photoelectron spectra and dichroism to previously reported XMCD
spectra of Fe containing Heusler
compounds~\cite{Elmers03,*Elmers04,Felser03,Wur05,*Wur06,*Wur2006} were the XMCD
spectra and dichroism appear rather without any resolved splitting of the
$L_{2,3}$ lines.

%%%%%%%%%%%%%%%%%%%%%%%%%%%%%%%%%%%%%%%%%%%%%%%%%%%%%%%%
\section{Summary and conclusions}
%%%%%%%%%%%%%%%%%%%%%%%%%%%%%%%%%%%%%%%%%%%%%%%%%%%%%%%%

In summary, MCDAD in hard X-ray photoelectron spectroscopy was used to study the
magnetic response of the core level of buried, remanently magnetized layers.
Using bulk-sensitive HAXPES-MCDAD, it was shown that IrMn exchange-biasing
layers keep thin films of CoFe or Co$_2$FeAl remanently magnetized in a well-
defined direction. Dichroism in the valence band spectroscopy is complicated in
metal/metal layers; however, the situation will improve in metal/insulator
structures in which the insulator does not contribute to the states at the Fermi
energy~\cite{FBG08}. The magnetic dichroism from core levels, including shallow
core levels, of CoFe and buried Co$_2$FeAl multilayer has asymmetries up to
above 58\% when it is excited by circularly polarized hard X-rays and is thus
much larger as compared to that in the case of excitation by soft X-rays. As a
noteworthy result, the differences in the Fe $2p$ emission from a regular
ferromagnet (CoFe) and a suggested half-metallic ferromagnet (Co$_2$FeAl) were
demonstrated. The splitting observed in Co$_2$FeAl points on the covalent
character of the compound.

Overall, the high bulk sensitivity of HAXPES combined with circularly polarized
photons will have a major impact on the study of the magnetic phenomena of
deeply buried magnetic materials. The combination with recently proposed
standing wave methods~\cite{Fadley10,Zeg10} will allow an element-specific study
of the magnetism of buried layers and make feasible the investigation of the
properties of magnetic layers not only at the surface but also at buried
interfaces.

%%%%%%%%%%%%%%%%%%%%%%%%%%%%%%%%%%%%%%%%%%%%%%%%%%%%%%%%%%%%%%%%%%%%%%
\bigskip
\begin{acknowledgments}

Financial support by Deutsche Forschungsgemeinschaft and the Strategic
International Cooperative Program of JST (DFG-JST:
FE633/6-1) is gratefully acknowledged. We are thankful to
the Japan Synchrotron Radiation Research Institute (JASRI) for the support
of experiments within approved 2009B0017 proposal. The team of Hokkaido University
acknowledges the support of MEXT, Japan (Grants-in-Aid 20246054, 21360140 and 19048001)
X.K. acknowledges the support of the graduate school of excellence MAINZ.

\end{acknowledgments}

\bibliography{mcdad_2}

\end{document}